\documentclass[letter]{aa} 
\pdfoutput=1
\usepackage{txfonts}
\usepackage{graphicx}
\usepackage{natbib}
\newcommand{\FIG}[1]{#1}
\bibpunct{(}{)}{;}{a}{}{,}
\begin{document}

\title{GRB blastwaves through wind-shaped circumburst media}
\titlerunning{GRB blastwaves through winds}
\offprints{Z. Meliani}
 \author{Z. Meliani$^1$, R. Keppens$^{2,1,3}$}
\institute{
$^1$ FOM-Institute for Plasma Physics Rijnhuizen P.O. Box 1207 3430 BE Nieuwegein, The Netherlands\\
$^2$ Centre for Plasma Astrophysics, K.U.Leuven, Belgium\\
$^3$ Astronomical Institute, Utrecht University, The Netherlands\\
\email{meliani@rijnh.nl, Rony.Keppens@wis.kuleuven.be}}

\date{Accepted;  Received }

\abstract{
A significant fraction of progenitors for long gamma-ray bursts
(GRBs) are believed to be massive stars. The investigation of long
GRBs therefore requires  modeling the propagation of
ultra-relativistic blastwaves through the circumburst medium
surrounding massive stars. We simulate the expansion of an
isotropic, adiabatic relativistic fireball into the wind-shaped
medium around a massive GRB progenitor. The
circumburst medium is composed of a realistically stratified stellar
wind zone up to its termination shock, followed by a region of
shocked wind characterized by a constant density.}
{
We followed the
evolution of the blastwave through all its stages, including the extremely rapid acceleration up to a Lorentz factor 75 flow, its deceleration
by interaction with stellar wind, its passage of the wind termination shock, until its propagation through shocked wind.}
{
We used the adaptive mesh refinement versatile advection code to follow the evolution of the fireball, from 3.3 seconds after its
initial release up to more than 4.5 days beyond the burst.}
{
We show that the acceleration from purely thermal to ultra-relativistic kinetic regimes is abrupt and produces an internally structured blastwave.
We resolved the structure of this ultra-relativistic shell in all stages, thanks to the adaptive mesh. We comment on the dynamical roles played by forward and
reverse shock pairs in the phase of interaction with the free stellar wind
and clearly identify  the complex shock-dominated structure created when the shell crosses the
terminal shock.}
{
 We show that in our model where the terminal shock is
taken relatively close to the massive star, the phase of self-similar deceleration of Blandford-McKee type can only be produced in the constant-density, shocked wind zone.}
\FIG{
}

\maketitle
\section{Motivation}
There is increasing evidence that the long-duration ($t_{\rm GRB}> 2
{\rm s}$) gamma-ray burst (GRB) is associated with the collapse of a
massive star with $M \ge 20 {\rm M}_{\odot}$ \citep{Larsonetal07}.
This evidence is supported by the association of some GRBs with a
supernova \citep{Galamaetal98, Woosley&Bloom06}, and also by the
association of GRBs with massive star-forming regions in distant
galaxies \citep{Wijersetal98, Trenthametal02}. { As known for
supernova blastwave modeling, the surroundings of the exploding
stars can influence its propagation.} Furthermore, some radio and
optical observations are consistent with a scenario of GRB ejecta
expanding into a CircumBurst Medium (CBM) with a wind density
profile $\rho\propto r^{-2}$ \citep{Panaitescu&Kumar04}. { Since
massive stars have significant mass-loss rates and structured
wind-blown bubbles surrounding their wind zone, we here investigate
the ejecta dynamics as it propagates through the bubble.}

Recently, significant progress has been made in investigating the dynamics of ultra-relativistic blastwaves
expanding in the CBM of a Wolf-Rayet (WR) star using analytical modeling \citep{Peer&Wijers06} and by numerical means, exploiting a Lagrangian relativistic hydro code \citep{Nakar&Granot06}.
We complement these efforts here by a numerical simulation of the complete fireball dynamics, expanding in the structured
CBM of a WR star.
We use grid-adaptive computations with AMRVAC
\citep{Keppens03} to investigate the ultra-relativistic hydrodynamic 
evolution of the fireball from its initial purely `hot' phase, up to times
significantly beyond its interaction with the transition from
free-wind to shocked-wind zones. To make a grid-adaptive 
computation even feasible, this wind termination shock is supposed to
 already occur at $R\sim 10^{16} {\rm cm}$, which is close to the
progenitor compared to values given by models of WR evolution by
\citet{Castoretal75}. Since the radius of a WR star is within the order
$3-11 R_\odot$
\citep{Meynetetal06}, this still turns  into a need to resolve
ultra-relativistic blastwave dynamics over a distance of at least 6
orders of magnitude. \citet{VanMarleetal06} explored a number of
physical mechanisms that could explain a more restricted free-wind
region, such as a high interstellar density and/or pressure, or a
lower mass loss rate of the WR star. One of the aims of this paper
is to quantify the effect of a sudden, termination-shock variation
in the CBM density profile on the dynamics of the blastwave. Some
scenarios suggest  reproducing the peculiar light-curve evolution of
some GRBs (e.g. 990123, 021211, 050904) \citep{Panaitescu&Kumar04,
Gendreetal07} by invoking an encounter of the blastwave with the
density jump across the wind termination shock. This can lead to a
brief brightening of the afterglow \citep{Wijers01}. In such a
scenario, the fireball expands during the first hours in a free-wind
medium, and after several hours (or few days), the blast further
decelerates in shock-dominated interactions with the constant
density medium representing shocked wind \citep{Gendreetal07}. {
Before modeling the effects of such an encounter on the spectral
changes in the light curve, we here model the high energy dynamics
associated with this terminal shock encounter. We investigate this
for the first time numerically for blastwaves in CBM of massive
stars.}
\begin{figure}
\begin{center}
\FIG{
{\rotatebox{0}{\resizebox{0.9\columnwidth}{4cm}{\includegraphics{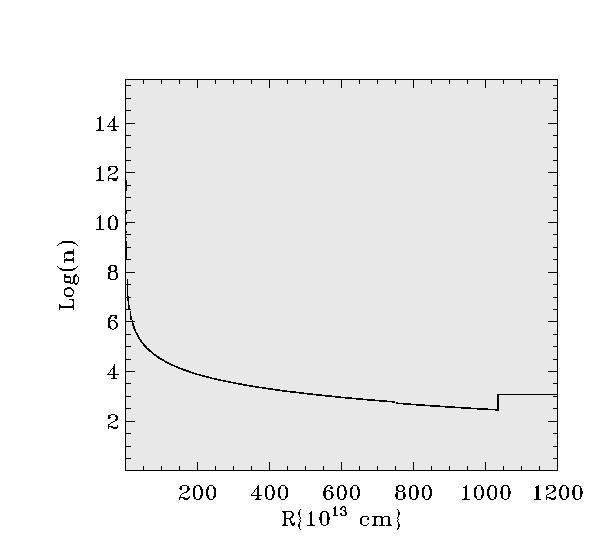}}}}
} \caption{The density profile of the CBM in our model.}
\label{LabelFig1}
\end{center}
\vspace{-1.0cm}
\end{figure}
\vspace{-0.5cm}

\section{The circumburst medium}
Wolf-Rayet stars are possible progenitors  GRB. During the WR phase, 
a massive star with mass $M \ge 20 {\rm M}_{\odot}$ has a
mass-loss rate of $\dot{M}_{\rm wind}=10^{-6}{\rm \, to\,} 10^{-4} {\rm
M}_{\odot}/{\rm yr}$ and a fast supersonic wind with speeds of $v_{\rm wind} =
1000 \;{\rm to}\; 2500 {\rm km/s}$ \citep{Chiosi&Maeder86}. This
wind interacts strongly with the surrounding medium, creating
two shocks. At the forward shock, the ambient medium is swept up, where it gets
compressed and heated. The reverse shock decelerates the
wind itself, and converts almost all the kinetic energy of the wind
to thermal energy, producing a hot gas with sound speed on the order
of the free-wind speed \citep{Weaveretal77}. The two shocked regions
are separated by a contact discontinuity, where Rayleigh-Taylor instabilities develop and lead to mixing. The
shocked ISM is very dense and cools quickly \citep{Francoetal94}.
However, the density in the shocked wind zone is lower, so its
cooling time is longer. The result
is an innermost zone with a hypersonic stellar wind and a second hot and almost
isobaric region. It consists of shocked stellar wind, mixed with
swept-up ISM. In fact, the total mass in this shocked wind region alone is
dominated by mass mixed in and evaporated from the shocked ISM ahead of the contact discontinuity. Still, most swept-up ISM is concentrated in a third region between contact discontinuity and forward shock, where it forms a
thin, dense, cold shell. The mass in this thin shell remains dominant throughout the stellar-wind bubble evolution \citep{Weaveretal77}.
A fourth region is the undisturbed interstellar medium.

The density profile in the free stellar wind is set to $n_{\rm wind}(r)\,=\,\dot{M}_{\rm wind}/(4\,\pi v_{\rm wind} r^2 m_{\rm p})$.
Here,
$m_{\rm p}$ is the mass of the proton.
The shocked wind region is isobaric as the sound speed in this region is higher than the
speed of expansion of the bubble, i.e. higher than
the speed of the forward shock \citep{Castoretal75, Weaveretal77}. In this
case, the swept-up shell
of ambient medium is driven by the pressure of the shocked wind. The pressure in this shocked wind region is then calculated at time $t$ from
\begin{eqnarray}\label{PRSW}
p_{\rm eq, b} &=& \frac{7}{(3850\,\pi)^{2/5}}
\left(\frac{\dot{M}_{\rm wind}\,v_{\rm wind}^{2}\rho_{\rm ISM}^{3/2}}{t^{2}}\right)^{2/5} \,,
\end{eqnarray}
and the radius of the bubble is computed from
\begin{eqnarray}\label{RFSW}
R_{\rm FS, wind} &=& \left(\frac{250}{308\pi}\right)^{1/5}
\left(\frac{\dot{M}_{\rm wind}\,v_{\rm wind}^{2}\,t^{3}}{\rho_{\rm ISM}}\right)^{1/5} \,.
\end{eqnarray}
In these expressions, we introduced
the density of the surrounding ISM $\rho_{\rm ISM}$.

The transition from free to shocked wind is at the location of the reverse shock, and
this is given by the balance between the
thermal pressure in the shocked wind region and the wind ram pressure (true 
in the case
of the energy driven phase when the shocked wind region cools slowly).
The strong shock condition at the reverse shock
$3 v_{\rm wind}^2 \rho_{\rm wind, eq}/4=p_{\rm eq}$ supposes that the pressure
in the shocked wind region is much higher than in the far free wind. This leads to
\begin{eqnarray}\label{RRSW}
R_{\rm RS, wind} &=& \left( \frac{3}{4}\frac{\dot{M}_{\rm wind}v_{\rm wind}}{4\pi p_{\rm eq, b}}\right)^{1/2}
\propto \left(\frac{\dot{M}_{\rm wind}^{3}v_{\rm wind}t^{4}}{\rho_{\rm ISM}^{3}}\right)^{1/10}\,.
\end{eqnarray}

In the following, we  model the fireball expanding  through only the free-wind and part of the shocked wind region.
Our wind termination shock will be at about $10^{16} {\rm cm}$, in accordance with Eq.~\ref{RRSW} for a high-density surrounding medium
with $\rho_{\rm ISM}=m_{\rm p} 10^5 {\rm cm}^{-3}$, a short WR lifetime
$t_{\rm WR}=100 {\rm year}$, 
a mass loss in the wind of $\dot{M}_{\rm wind} = 10^{-6}M_{\odot}/{\rm yr}$, and wind speed $v_{\rm wind}=10^3 {\rm km/s}$.
While the last two are typical values in some circumburst media for GRB
\citep{Arthur06},
we took a very short WR lifetime to
investigate the case where the terminal shock is at a much shorter distance  than the typical $1 {\rm pc}$ \citep{Chevalieretal04}.
The density profile of the CBM in our model is given
in Fig.\ref{LabelFig1}. The pressure profile in the free supersonic
stellar wind is deduced by considering a Mach number ${\cal M}=3$. The wind is then
cold, and its pressure will have no effect on the propagation of the fireball. In
the shocked wind, the constant pressure is given by Eq.\ref{PRSW}, while the constant density is increased fourfold in accordance with
the strong shock requirement. The wind velocity is constant throughout the wind zone and neglected in the shocked wind region.
This is a proxy for the reduction by a factor of 4 expected at the termination shock.
\vspace{-0.5cm}

\section{GRB shell model and evolution}
Initially we set a uniform static and hot shell
that extended up to radius $R_{\rm 0}=10^{11}{\rm cm}$.
The initial specific enthalpy, normalized to $c^2$, in the uniform shell is
$\bar{\eta}_{\rm sh}=100$, and its energy is
\begin{equation}
E_{\rm sh}=10^{51}{\rm ergs}=4\pi R_0^2 \delta \bar{\eta}_{\rm sh}c^2  n_{\rm sh} m_{\rm p}
\label{Eiso}
\end{equation}
where $\delta$ stands for the thickness of the shell. We set approximately
$\delta\sim c\,\Delta t \sim 10^{11}{\rm cm}$,
where $\Delta t\approx 3 {\rm s}$ is the duration of the GRB. The mass of the shell
is $M_{\rm sh} = E_{\rm sh}/(\bar{\eta}_{\rm sh}\,c^2)$.
The shell is static and hot, and the initial pressure is set to
$p_{\rm sh}=\frac{\Gamma-1}{\Gamma}n_{\rm sh}
\left(\bar{\eta}_{\rm sh}-1\right)\,m_{p}\,c^2 $ and the density  from Eq.~\ref{Eiso}.
Initially, the energy of the shell is only thermal. We use a constant polytropic
index $\Gamma=4/3$, as the temperature of the shell is relativistic and the
interaction shell-ISM will be dominated by the forward shock, where the temperature
of the shocked ISM is also relativistic. The computation is done on a domain extending from a radius
of $10^9 {\rm cm}$ to $1.2\times 10^{16} {\rm cm}$, with the initial shell region between
$2 \times 10^9 {\rm cm}$ and $10^{11} {\rm cm}$. In the grid-adaptive numerical simulation, we use 1200 grid points on the
base grid level, and allow for 15 grid levels in total, with a doubling of the effective resolution between
each grid level.

\begin{figure}
\begin{center}
\FIG{
{\rotatebox{0}{\resizebox{0.9\columnwidth}{3.8cm}{\includegraphics{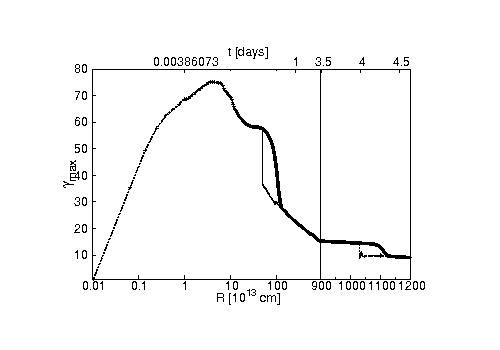}}}}
} \caption{The variation in the maximum
Lorentz factor. We also plot the Lorentz factor at the forward
shock. The variation is presented a function of the distance and in
the lab-frame time. Note the change from log to linear scale at $9
\times 10^{15} {\rm cm}$.} \label{LabelFig2}
\end{center}
\vspace{-0.5cm}
\end{figure}

\begin{figure}
\begin{center}
\FIG{
{\rotatebox{0}{\resizebox{0.9\columnwidth}{3.8cm}{\includegraphics{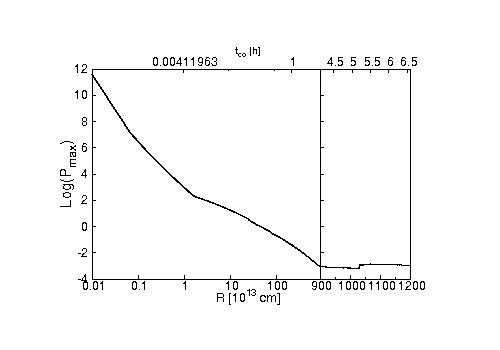}}}}
} \caption{The variation in the (logarithm of) maximum pressure as a
function of the distance and the comoving frame time.}
\label{LabelFig3}
\end{center}
\vspace{-0.5cm}
\end{figure}

In Fig.\ref{LabelFig2}, we draw the variation in time of the maximum
Lorentz factor of the fireball.
We also draw the variation of the
Lorentz factor at the forward shock alone: this coincides mostly with the instantaneous maximum Lorentz factor, except for those intervals where reverse shock dynamics is particularly prominent, as discussed below. In Fig.\ref{LabelFig3},
we draw the variation of the maximum pressure with time. Note that we translated to comoving (at the forward shock) time for the latter.

\begin{figure}
\begin{center}
\FIG{
{\rotatebox{0}{\resizebox{0.9\columnwidth}{2.6cm}{\includegraphics{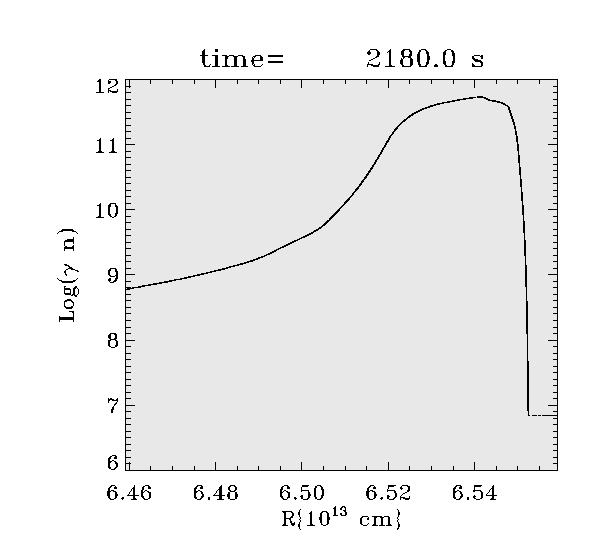}}}}
{\rotatebox{0}{\resizebox{0.9\columnwidth}{2.6cm}{\includegraphics{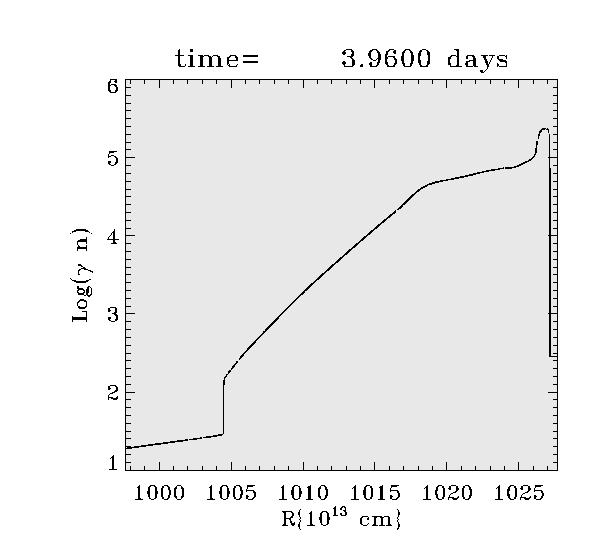}.jpg}}}
{\rotatebox{0}{\resizebox{0.9\columnwidth}{2.6cm}{\includegraphics{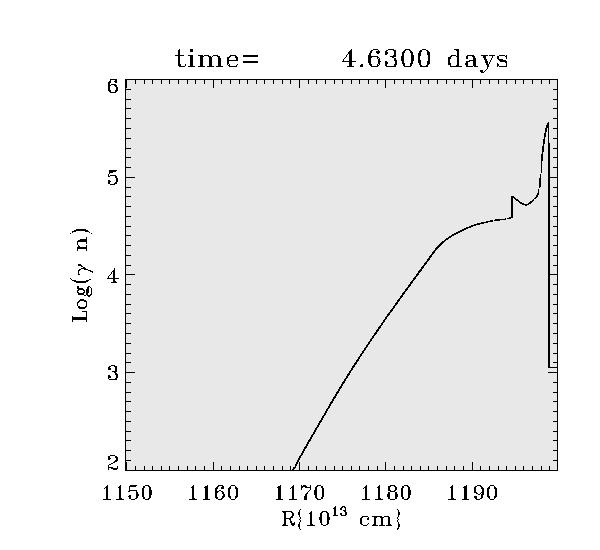}}}}
{\rotatebox{0}{\resizebox{0.9\columnwidth}{2.6cm}{\includegraphics{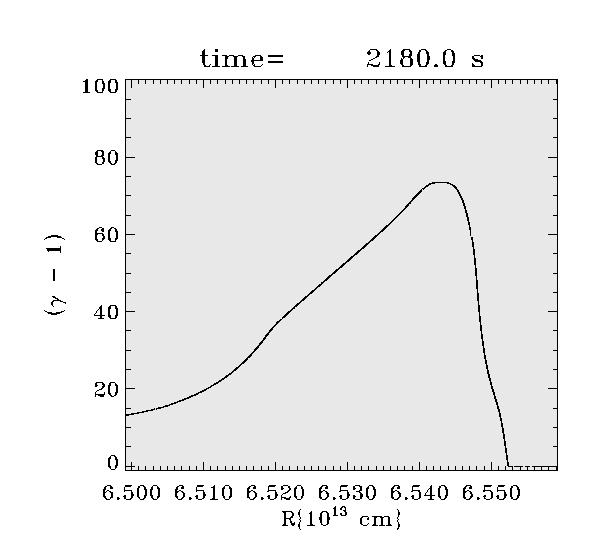}}}}
{\rotatebox{0}{\resizebox{0.9\columnwidth}{2.6cm}{\includegraphics{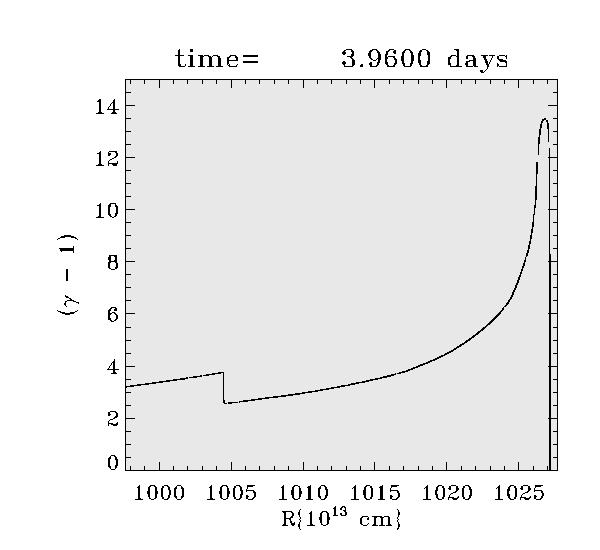}}}}
{\rotatebox{0}{\resizebox{0.9\columnwidth}{2.6cm}{\includegraphics{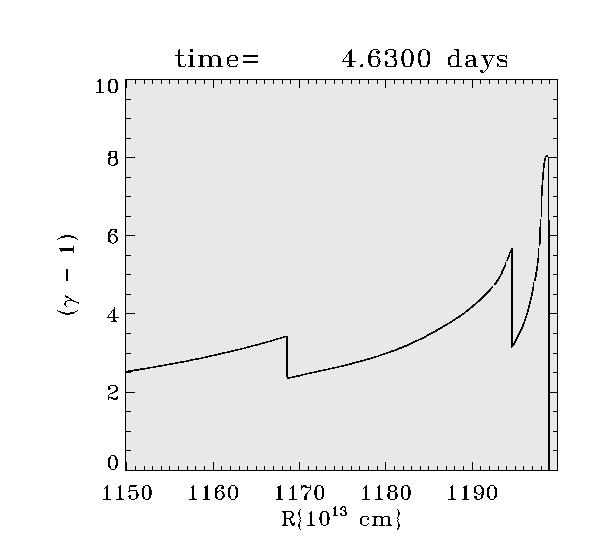}}}}
} \caption{All zones present when the relativistic shell interacts
with the CBM. The shell profile is shown at three different times: from top
down: at time $t=2180 {\rm s}$ after the rapid thermal acceleration;
at time $t=3.96 {\rm days}$, prior to the encounter with the wind
termination shock at time $t=4.63 {\rm days}$. The first three
frames show the logarithm of density in the lab-frame, the last three
show kinetic energy.} \label{LabelFig4}
\end{center}
\vspace{-1.0cm}
\end{figure}

In the first phase of the simulated fireball, the initial hot shell and a fraction of swept-up matter
are accelerated thermally (Fig.\ref{LabelFig3}) very fast, such that a maximal Lorentz factor of
$\gamma_{\rm max}=75$ is reached within $t_{\rm acc}\sim 1200 {\rm s}$ (Fig~\ref{LabelFig2} and \ref{LabelFig4}).
As the shell is initially
uniform, the center and back of the shell get delayed in their acceleration with respect to the front of the shell, and they reach a lower Lorentz factor and introduce a tail structure (Fig.~\ref{LabelFig4}).
The maximum Lorentz factor reached is low in the sense that $\gamma_{\rm max}<\bar{\eta}_{\rm sh}=100$, due to the influence of accreted
mass from the wind. This influence will generally depend on the initial energy in the shell
$E_{\rm sh}$ and the mass loss rate in the wind. In the simulation, in this acceleration phase
the shell has accreted and accelerates with it a mass of
$M_{\rm acc, wind}=0.024 M_{\rm sh}$.
If the realized energy in the GRB
is higher, the maximum Lorentz factor reached by the shell will increase, too.

After this rapid acceleration phase, the swept-up wind mass increases enough to
influence the dynamics of the shell. The shell now decelerates by transferring
kinetic energy at both a forward and a reverse shock pair.
An intermediate contact discontinuity separates shell from the swept-up, shocked wind
matter. At the forward shock, the kinetic energy of the shell is passed to a swept-up
shocked wind.
As the shell has an internal structure produced during the acceleration phase
(Fig.~\ref{LabelFig4}), the maximum Lorentz factor seen in the evolution
(Fig.~\ref{LabelFig2}) decreases smoothly. This is  contrasts with
the sudden change seen in simulations where a uniform shell travels through
constant medium density, as discussed in detail by \citet{Melianietal07}. In this more
realistic case, the reverse shock will cross that frontal part of the shell with the
highest Lorentz factor at about time
$t\sim 6\times 10^{3} {\rm s}$. In a second phase, the overall maximal Lorentz factor remains constant $\gamma\sim 58$ until
$t\sim 19 \times 10^{3} {\rm s}$. This coincides with those times when the reverse shock crosses the middle part of the shell,
which has acquired a constant Lorentz factor. 

After this, the reverse shock starts to propagate
in the tail of the shell and, in this phase, the reverse shock is mildly relativistic.
This lasts in Fig.\ref{LabelFig2} until $t\sim 9.4\,{\rm hours}$ marking a relatively fast decrease in the maximum Lorentz factor, which
actually coincides with the Lorentz factor in the, as yet, unshocked shell part ahead of the reverse shock. Beyond this
rapid phase, the maximum Lorentz factor seen is situated at the front of the forward shock. In this phase
more energy starts to be transferred to swept-up, shocked wind matter, and
the Lorentz factor at the forward shock decreases with time with a slope smaller than
$-1/2$. The latter dependence would be expected from by the analytic Blandford-McKee solution, but in our simulation not
all shell energy has already been transferred
to accreted matter, which is assumed in the self-similar solution.

When the shock-dominated shell structure finally reaches the wind termination
shock $R_{\rm RS, wind}$, all the matter originally in the free-wind zone
is swept-up by the shell and compressed between its forward shock and contact
discontinuity (Fig.~\ref{LabelFig4}).
As the forward shock is relativistic  (Fig.~\ref{LabelFig4}),
the compression ratio of the number density there is
$\frac{n_{\rm 2}}{n_{\rm 1}}\sim59.4\sim\left(4\gamma(r)+3\right)$,
where $n_2$ is the density in the compressed
wind matter, while $n_1$ is the original wind density. Similarly, the thermal
energy density in the compressed wind matter is
$e_{\rm 2}\sim\left(\gamma_{\rm 2}-1\right)\rho_{2} c^2$
\citep{Blandford&McKee76}, which corresponds to analytical estimates.
However, as stated above, the shell does not yet reach a self-similar Blandford-McKee phase while traversing the free-wind zone, as the
reverse shock continues to propagate in the tail of the shell and not all its
energy is transferred to swept-up matter. The Blandford-McKee phase can be
reached in a  free stellar-wind region when the terminal shock would be faraway.
Due to our assumption of a relatively close termination shock, our
numerical simulation allows us to investigate blast waves interacting with wind
termination shocks before the Blandford-McKee phase.

As soon as the shell structure has encountered the wind termination shock and
starts to travel through the shocked wind region,
the entire shell is actually composed of multiple regions
(Fig~\ref{LabelFig4}). From right to left, we find (1) a forward shock now
propagating in the shocked wind region.
We then encounter (2) a contact discontinuity separating the re-shocked shocked
wind from the swept-up free wind. This wind is shocked once more by
(3) a new reverse shock propagating through the already swept-up compressed wind
matter. This matter extends to (4) the original contact discontinuity between
shell and swept-up wind matter.
At this new reverse shock (3), the Lorentz factor
falls to $\gamma_{\rm  3}\sim 10 = 0.725 \gamma_{\rm  2}$ and the density
increases to $n_{\rm  3}\sim 28\times 10^{4}{\rm cm^{-3}} = \sqrt{3} n_{\rm  2}$.
Furthermore, the thermal energy $e_{\rm  3}\sim 2.1 e_{\rm  2}$ and this new
reverse shock remains Newtonian until the time $t=4.16 {\rm day}$.
The relations
between the values of the density and Lorentz factor are in good agreement with
recent analytical estimates by \cite{Peer&Wijers06}. Between (3), the new reverse
shock, and (4), the old contact discontinuity, there is at first still a region of
the swept-up wind matter with values for $n_2(r)$ (decreased due to spherical expansion) and $\gamma_2$ as discussed
above. Further leftward of (4) the old contact discontinuity, we find
all the initial shell material, separated by (5) the old reverse shock, which continues to propagate in the
tail of the initial shell. Hence, there are two regions in this part of the structure as well, namely shocked-shell matter
between the contact discontinuity (4) and the initial reverse
shock (5), and the original unshocked-shell matter.

The new reverse shock (3) eventually crosses all the region that extends up to the original contact discontinuity (4), and it arrives there
at a distance $R_{1}\sim 1.1 \times 10^{16} {\rm cm}$.
For times up to this arrival at
$R_{1}$, the maximum Lorentz factor seen in Fig.\ref{LabelFig2} undergoes a weak
variation, since the Lorentz factor $\gamma_2$, density $n_2$, and
energy in the swept-up wind region vary little. After arrival at $R_{1}$, the maximum Lorentz factor drawn in
Fig.\ref{LabelFig2} represents the value in the region of shocked-shell matter ahead
of the new reverse (Newtonian) shock, where
the Lorentz factor decreases more strongly. After $R>1130 \times 10^{13}{\rm cm}$, the
maximum Lorentz factor drawn in Fig.\ref{LabelFig2} starts to represent the value at the
forward shock (1) again, in the phase before it reaches the Blandford-McKee phase.
 From then on, while the new
reverse shock continues to propagate through the original shell structure, the maximum Lorentz factor
at the forward shock
starts to decrease with distance with a slope $-0.9$. This is still less than
the slope in the expected long-term Blandford-McKee phase. To arrive at this phase, a still larger computational domain and a larger simulation time is needed.

In Fig.~\ref{LabelFig2}, the time delay in the drop of the maximum
Lorentz factor from when the shell crosses the wind terminal shock
is thus related to the time for the new reverse shock to propagate
through the region (2)-(4). Note that the maximal pressure in
Fig.~\ref{LabelFig3} directly increases at the wind terminal shock
encounter. \vspace{-0.5cm}

\section{Conclusions}
\vspace{-0.2cm}

 In this work, we have investigated all phases of the GRB
in a fireball modeled. We model the propagation of the thermal
fireball through a complex wind-shaped CBM of a massive star. The
fireball interacts with free stellar wind, crosses the wind
terminal shock, and then propagates in the shocked wind zone. We
discussed initial acceleration, energy transfer to CBM,
deceleration, and interaction with the terminal shock. When the
shell reaches the wind termination shock, the structure of the shell
changes and multiple regions form: a forward shock, two contact
discontinuities, and two reverse shocks characterize the evolving
structure. We have shown that our simulations agree in terms of
compression ratios, Lorentz factors, and energies reached at all
these shocks with analytical estimates exploiting the jump
conditions \citep{Peer&Wijers06}.

There are pronounced differences in the deceleration as quantified
by the variation in the instantaneous maximum Lorentz factor with
models of the afterglow phase alone \citep{Melianietal07}. We showed
that the rapid thermal acceleration produces a cold shell with
internal structure where the Lorentz factor and density decrease
from the head to the back of the shell. This has distinct
consequences for its further long-term evolution and deceleration.
In the deceleration phase before Blandford-McKee, the shell
decelerates with a slope $-0.9$ in the constant density medium. 
These results obtained using high-resolution numerical simulation
bring out important differences with analytical estimates, in
particular with respect to the internal structure of the expanding
high-energy shell. In future work, the results obtained with our
model will be used to deduce light curves, and to show how sudden
rebrightenings may help to deduce the position of the terminal
shock. We have shown here that it is not
appropriate to use the self-similar Blandford-McKee solution  around the terminal shock.
 Future work will explore multi-dimensional scenarios and
quantify the spectral outcome.
\vspace{-0.5cm}

\section*{Acknowledgements}
\vspace{-0.2cm}

We acknowledge financial support from NWO-E grant 614.000.421 and
computing resources by NCF grant SG-06-276 and the VIC cluster at
K.U.Leuven.

\end{document}